\documentclass[aps,pra,twocolumn,superscriptaddress,10pt,article,nofootinbib,showpacs]{revtex4-1}
\usepackage{amsfonts}
\usepackage{amsmath}
\usepackage{amsthm}
\usepackage{braket}
\usepackage{graphicx}
\usepackage{hyperref}
\usepackage{color}
\usepackage{amssymb}
\usepackage{dsfont}
\usepackage{verbatim}
\usepackage{amsmath,amscd}
\usepackage{multirow}

\newcommand{\mathscr}[1]{\ensuremath{\mathcal{#1}}}

\begin{document}

\title{Global anomaly detection in two-dimensional symmetry-protected topological phases}

\author{Nick Bultinck}
\affiliation{Department of Physics and Astronomy, Ghent University, Krijgslaan 281 S9, B-9000 Ghent, Belgium}
\author{Robijn Vanhove}
\affiliation{Department of Physics and Astronomy, Ghent University, Krijgslaan 281 S9, B-9000 Ghent, Belgium}
\author{Jutho Haegeman}
\affiliation{Department of Physics and Astronomy, Ghent University, Krijgslaan 281 S9, B-9000 Ghent, Belgium}
\author{Frank Verstraete}
\affiliation{Department of Physics and Astronomy, Ghent University, Krijgslaan 281 S9, B-9000 Ghent, Belgium}
\affiliation{Vienna Center for Quantum Technology, University of Vienna, Boltzmanngasse
5, 1090 Vienna, Austria}

\begin{abstract}
Edge theories of symmetry-protected topological phases are well-known to possess global symmetry anomalies. In this work we focus on two-dimensional bosonic phases protected by an on-site symmetry and analyse the corresponding edge anomalies in more detail. Physical interpretations of the anomaly in terms of an obstruction to orbifolding and constructing symmetry-preserving boundaries are connected to the cohomology classification of symmetry-protected phases in two dimensions. Using the tensor network and matrix product state formalism we numerically illustrate our arguments and discuss computational detection schemes to identify symmetry-protected order in a ground state wave function.
\end{abstract}

\maketitle

Global symmetries of local quantum many-body Hamiltonians can be implemented in fundamentally different ways in the corresponding low-energy states, even if there is no spontaneous symmetry breaking. This observation has led to the concept of symmetry-protected (and symmetry-enriched) topological phases \cite{ChenLiu,ChenGu,PollmannBerg,FidkowskiKitaev}, which have been the subject of intense research in recent years. The different low-energy symmetry realizations were found to manifest themselves in various physical properties. For instance, symmetry defects bind fractional charges and with open boundaries, Symmetry Protected Topological (SPT) phases admit edge degrees of freedom which cannot exist without the presence of the higher-dimensional bulk. It was subsequently realized that these edge theories possess a fatal global symmetry anomaly which prevents them from being realizable as an independent lattice system \cite{Wen,KapustinThorngren,WangSantos1}.  

In this work we focus on SPT phases in two spatial dimensions. We also restrict ourselves to bosonic systems with discrete, unitary on-site symmetries. SPT phases are characterized by an energy gap and a unique ground state when defined with closed boundary conditions, implying that there are no non-trivial topological superselection sectors. They are adiabatically connected to the trivial product state phase when the global symmetries are allowed to be broken. In two dimensions, the edge of a SPT phase has to either break the symmetry or be gapless. We will only consider the gapless case here.

Our goal in this work is two-sided. First, we want to deepen the understanding of anomalies associated to unitary global symmetries in one dimensional gapless systems, and their connection to SPT phases in two dimensions. Secondly, we want to use the anomaly to study SPT phases numerically. For our numerical results we make great use of the Matrix Product Operator (MPO) formalism for SPT phases \cite{ChenLiu,SPTpaper}. All SPT ground states we use are tensor networks where the virtual symmetry action is implemented by a MPO. These MPOs are explicit lattice realizations of the anomalous symmetry action on the edge \cite{ChenLiu}, and the group defect lines we use in our theoretical arguments. For related work on lattice defects and the connection between tensor network methods and Conformal Field Theories (CFTs), see \cite{ChuiMercat,Aasen,Bridgeman1,BridgemanOBrien,HauruEvenbly,Bal}.

Below we first discuss how group cohomology, underlying the SPT classification \cite{ChenGu}, arises in CFT via group defect lines. We then establish a relation between group cohomology and an obstruction to both orbifolding \cite{RyuZhang,SuleChen,HsiehSule} and finding symmetric boundary conditions \cite{HanTiwari}. We numerically illustrate the appearance of non-trivial cohomology classes in CFTs associated to two-dimensional SPT phases using the strange correlator method~\cite{YouBi}. At the end, we discuss how CFT techniques can be exploited to optimize numerical detection of SPT phases via the entanglement spectrum. Numerical schemes to uniquely determine a two-dimensional SPT phase already exist \cite{Zaletel}, but do not rely on the entanglement spectrum.

\begin{figure}[t]
a)
\includegraphics[width=0.075\textwidth]{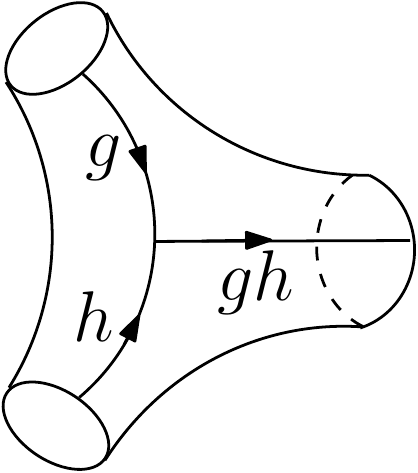}\hspace{4 mm} b)
\includegraphics[width=0.33\textwidth]{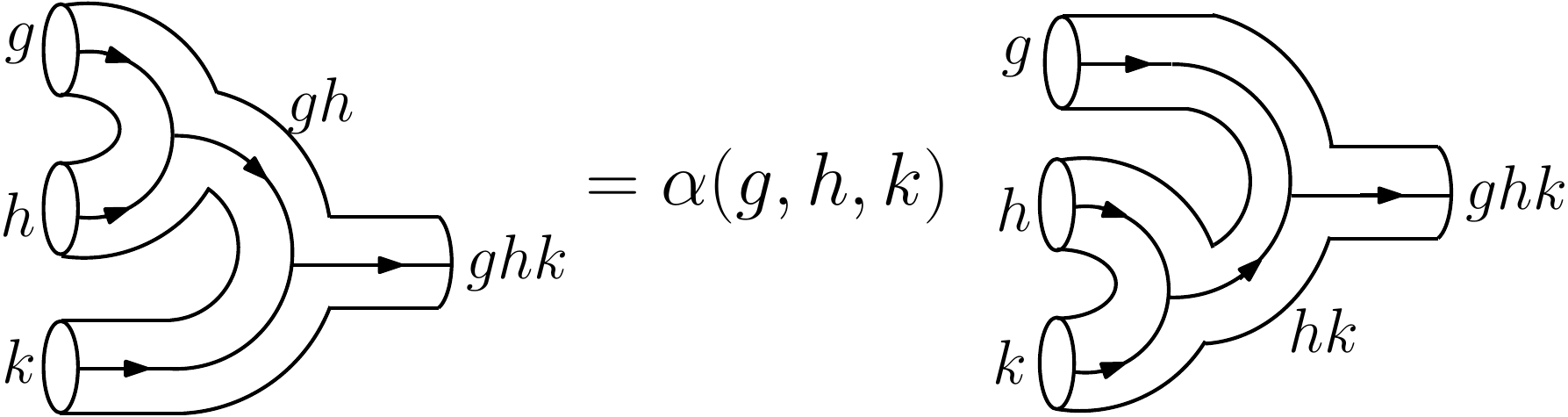}
\caption{(a) Vertex operator with twisted boundary conditions labeled by group elements $g,h$ and $gh$. (b) Operators associated to four-point functions of twist fields can be decomposed in two different ways in vertex operators.}\label{fig:vertex}
\end{figure}

\emph{Group defect lines -- } We use conformal field theory to describe the gapless one-dimensional systems localized at the edge of the two-dimensional SPT bulk. By assumption, these CFTs inherit the global symmetry group $G$ from the bulk, which provides the possibility of introducing twist fields $\sigma^i_g(z)$ \cite{DixonFriedan}. A twist field $\sigma^i_g(z)$ has a branch cut attached to it such that any field crossing the cut changes with the group action corresponding to $g\in G$. The branch cuts associated to the twist fields are special instances of topological defect lines \cite{Oshikawa,Petkova,Bachas}, which in the case of Rational Conformal Field Theories (RCFT) are well-understood \cite{FrohlichFuchs2,FuchsGaberdiel}. We will therefore refer to these branch cuts as group defect lines, which play a central role in this manuscript. Let us now consider the vertex operator $V_{(g,h)}$, which is represented by the path integral on a pair-of-pants-manifold with twisted boundary conditions as shown in Figure \ref{fig:vertex}(a). Note that via the operator-state correspondance $V_{(g,h)}$ contains information about three-point functions of twist fields. In our graphical notation, group defect lines are equipped with arrows, which indicate what direction of crossing the line corresponds to an action of $g$ and what direction to an action of $g^{-1}$. Reversing an arrow is equivalent to changing the label $g$ by $g^{-1}$. We also pick the convention that the arrows on defect lines indicate the orientation of the boundary on which they terminate. We can multiply every vertex operator $V_{(g,h)}$ with a different phase $\beta(g,h)$, with the restriction that $\beta(e,g)=\beta(g,e)=1$ for all $g \in G$, where $e$ is the identity group element. Once we fix these phases, the phase of the path integral on other manifolds is also automatically fixed by cutting and gluing. Phases of the form $b(g)b(h)b(gh)^{-1}$ are trivial since they can be absorbed in the states on the boundary of the vertex operator. 

The operator associated to four-point functions of twist fields can be decomposed in two different ways in terms of vertex operators $V_{(g,h)}$ as illustrated in Figure \ref{fig:vertex}(b). These two different decompositions differ by a phase $\alpha(g,h,k)$, which satisfies $\alpha(e,g,h)=\alpha(g,e,h)=\alpha(g,h,e)=1$ for all $g,h \in G$. By considering operators associated to five-point functions one can show that this phase must satisfy the consistency condition $\alpha(g,h,k)\alpha(g,hk,l)\alpha(h,k,l)=\alpha(gh,k,l)\alpha(g,h,kl)$, implying that $\alpha(g,h,k)$ is a group 3-cocycle \cite{MooreSeiberg}. Assigning different phases $\beta(g,h)$ to the vertex operators $V_{(g,h)}$ changes the 3-cocycle $\alpha(g,h,k)$ by a coboundary to $\alpha(g,h,k)\beta(g,h)\beta(gh,k)\beta(g,hk)^{-1}\beta(h,k)^{-1}$. This defines an equivalence relation on solutions of the 3-cocycle relation. The different equivalence classes constitute the third cohomology group $H^3(G,U(1))$.

\emph{Orbifolding and boundary conditions -- } In Ref. \cite{RyuZhang,SuleChen,HsiehSule} it was observed that orbifolds of SPT edge theories cannot be made modular invariant. We will argue that for CFTs with $\alpha(g,h,k)$ in a non-trivial cohomology class there is indeed a conflict between orbifolding and global diffeomorphism invariance. For RCFT, it is already known that $H^3(G,U(1))$ forms an obstruction to orbifolding \cite{FrohlichFuchs1,FuchsRunkel}. Our argument holds for both rational and irrational CFTs. In the path integral formalism, the orbifold partition function on a torus is obtained by summing over partition functions with boundary conditions along the two non-contractible cycles twisted by group elements $g$ and $h$ \cite{DixonHarvey,Dijkgraaf}. To get a consistent twisted partition function $g$ and $h$ have to commute. The twisted boundary conditions are implemented by inserting appropriate group defect lines. The crossing point of the defect lines has to be resolved using trivalent junctions as those appearing in the vertex operators $V_{(g,h)}$. The torus partition function of the orbifold theory is thus

\begin{align}
Z_{\text{orb}}(\tau)=\sum_{g,h\in G\, |\,[g,h]=e} \epsilon(g,h)
 & \hspace{2 mm} \raisebox{-.4\height}{\includegraphics[scale=0.32]{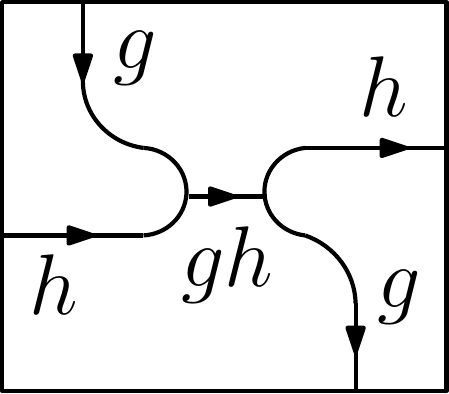}}\; ,
\end{align}

where we represent the torus as a square with opposite sides identified. $\tau$ is the modular parameter of the torus and $[g,h]=ghg^{-1}h^{-1}$. The phases $\epsilon(g,h)$ occuring in this sum are called discrete torsion~\cite{Vafa1} and are given by $\epsilon(g,h)=\beta(g,h)\beta(h,g)^{-1}$.

Using the results of appendix D and E of Ref.~\cite{SPTpaper} one can derive the transformation of the orbifold partition function under the modular group. Under a Dehn twist, or $T$ transformation, the partition function becomes

\begin{align}\label{eq:T}
Z_{\text{orb}}(\tau +1)=\sum_{g,h\in G\, |\,[g,h]=e} \frac{\epsilon(g,h)}{\alpha(h,g,h)}
 & \hspace{2 mm} \raisebox{-.4\height}{\includegraphics[scale=0.32]{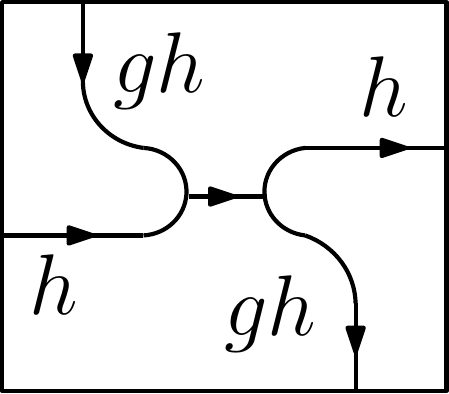}}\; ,
\end{align}
while under the $S$ transformation we get

\begin{align}\label{eq:S}
Z_{\text{orb}}(-1/\tau)=\sum_{g,h\in G\, |\,[g,h]=e} \epsilon(g,h)\omega_g(h,h^{-1})
 & \hspace{2 mm} \raisebox{-.4\height}{\includegraphics[scale=0.32]{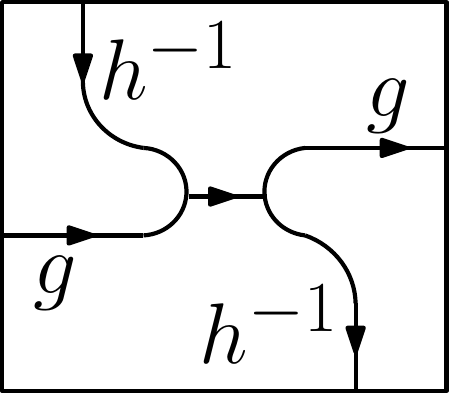}}\; ,
\end{align}
where $\omega_g(h,h^{-1})=\alpha(g,h,h^{-1})\alpha(h,h^{-1},g)\alpha(h,g,h^{-1})^{-1}$. We derive these expressions explicitly in the supplementary material. Let us first consider the case $\alpha \equiv 1$. In Ref.~\cite{Vafa1} a set of conditions on $\epsilon(g,h)$ were derived based on modular invariance of the torus and higher genus partition functions. One can check that these conditions are equivalent to the requirement that $\beta(g,h)$ is a 2-cocycle, i.e. $\beta(g,h)$ satisfies $\beta(g,h)\beta(gh,k)=\beta(g,hk)\beta(h,k)$. This produces the known result that the number of different orbifolds corresponds to the number of second cohomology classes $H^2(G,U(1))$. When $\alpha\neq 1$, equations \eqref{eq:T} and \eqref{eq:S} are equivalent to the expressions derived previously for the $T$ and $S$ matrices in the ground state subspace of the corresponding two-dimensional Dijkgraaf-Witten gauge theory on the torus \cite{DijkgraafWitten,HuWan}. Thus modular invariance implies that $\alpha(g,h,k)$ belongs to the trivial cohomology class in $H^3(G,U(1))$, provided that all twisted Dijkgraaf-Witten gauge theories can be distinguished from the untwisted gauge theory via these $T$ and $S$ matrices. For 3-cocycles where this is not possible --see Ref.~\cite{Mignard}-- we suspect that the anomaly can only be detected on higher-genus Riemann surfaces.

\begin{figure}
a) 
\includegraphics[width=0.21\textwidth]{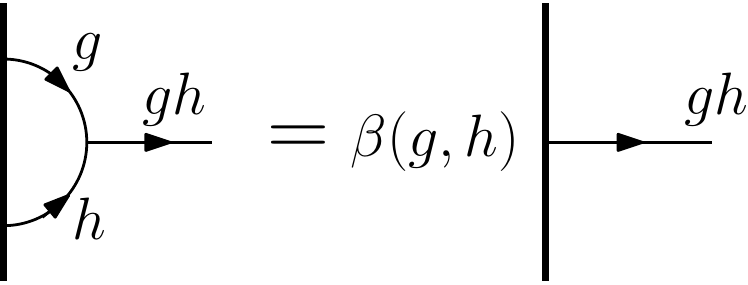}\hspace{2 mm} b)
\includegraphics[width=0.21\textwidth]{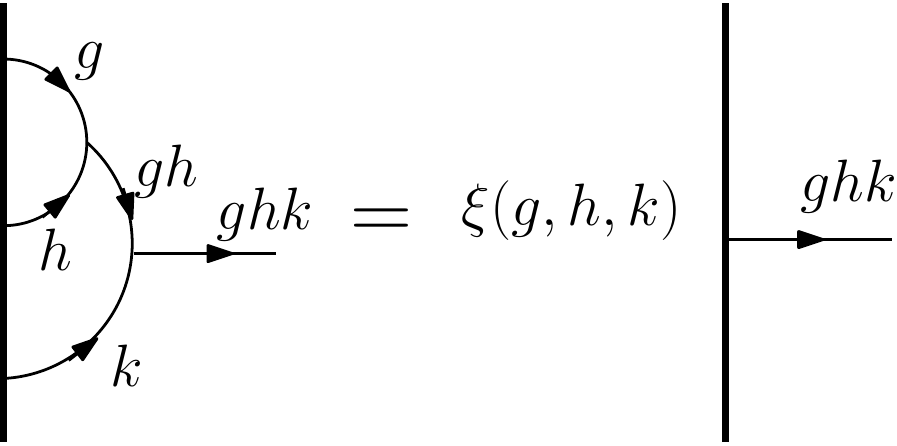}
\caption{Properties of symmetry preserving boundary conditions. Fat vertical lines represent the CFT boundary, lines with arrows are group defect lines.}\label{fig:boundary}
\end{figure}

In Ref.~\cite{HanTiwari} it was argued that anomalous edge theories of two-dimensional SPT phases do not admit boundary conditions preserving both the conformal symmetry and the global symmetry $G$. It is easy to show that for CFTs with a non-trivial 3-cocycle $\alpha(g,h,k)$ such boundary conditions indeed do not exist. To see this, one simply needs to realize that a global symmetry action on the Hilbert space of the space-like open interval is represented in the path integral by a group defect line going from one time-like boundary to the other. Conformal boundary conditions preserving the global symmetry $G$ should then satisfy the condition shown in Figure \ref{fig:boundary}(a). Now looking at Figure \ref{fig:boundary}(b), the phase $\xi(g,h,k)$ depicted there can be derived in two different ways; equality of both phases then implies $\alpha(g,h,k)=\beta(g,h)\beta(gh,k)\beta(g,hk)^{-1}\beta(h,k)^{-1}$, from which it follows that $\alpha(g,h,k)$ indeed belongs to the trivial class in $H^3(G,U(1))$.

\begin{figure}[t]
a)
\includegraphics[width=0.2\textwidth]{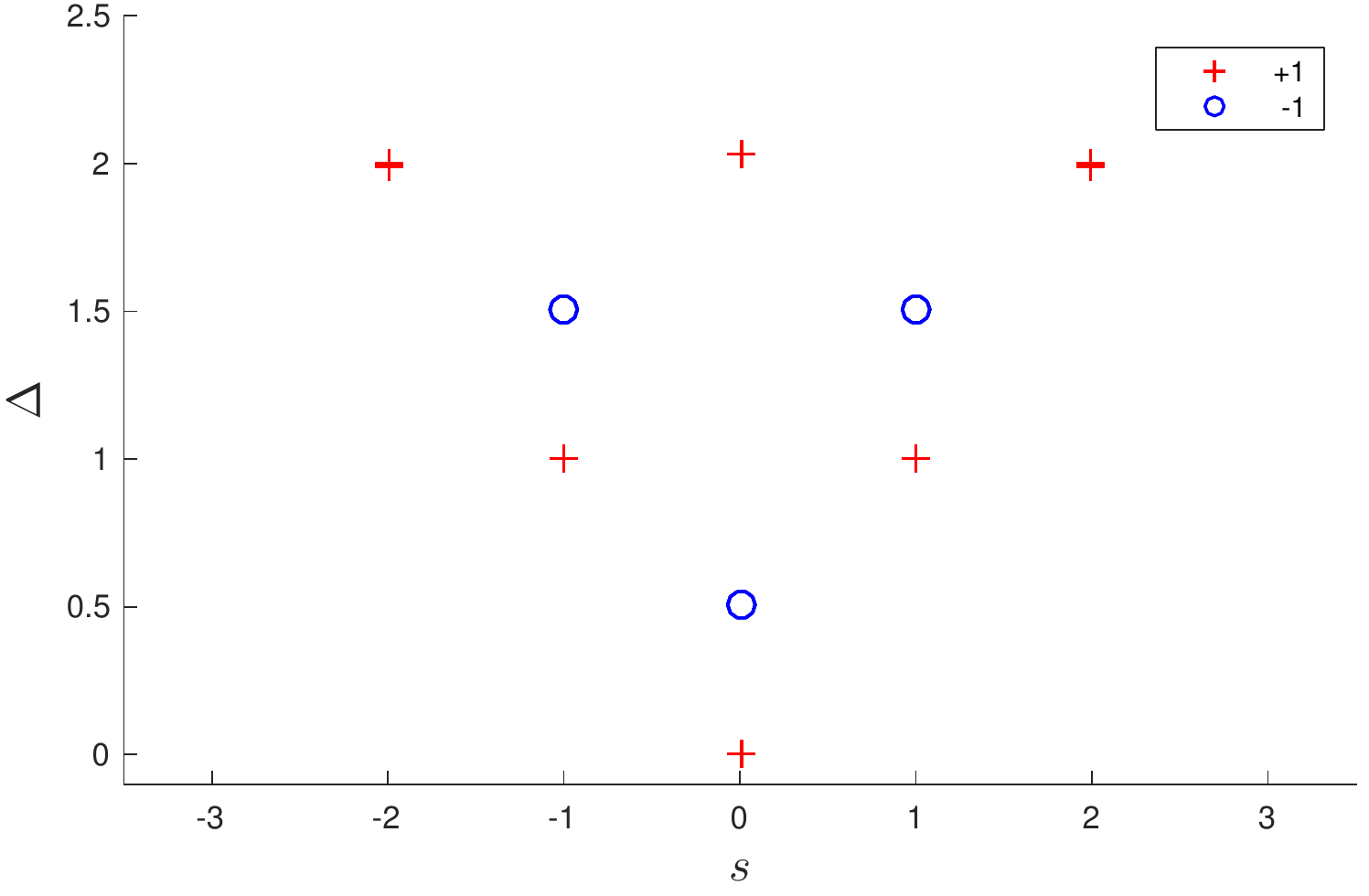} \hspace{1 mm}b)
\includegraphics[width=0.2\textwidth]{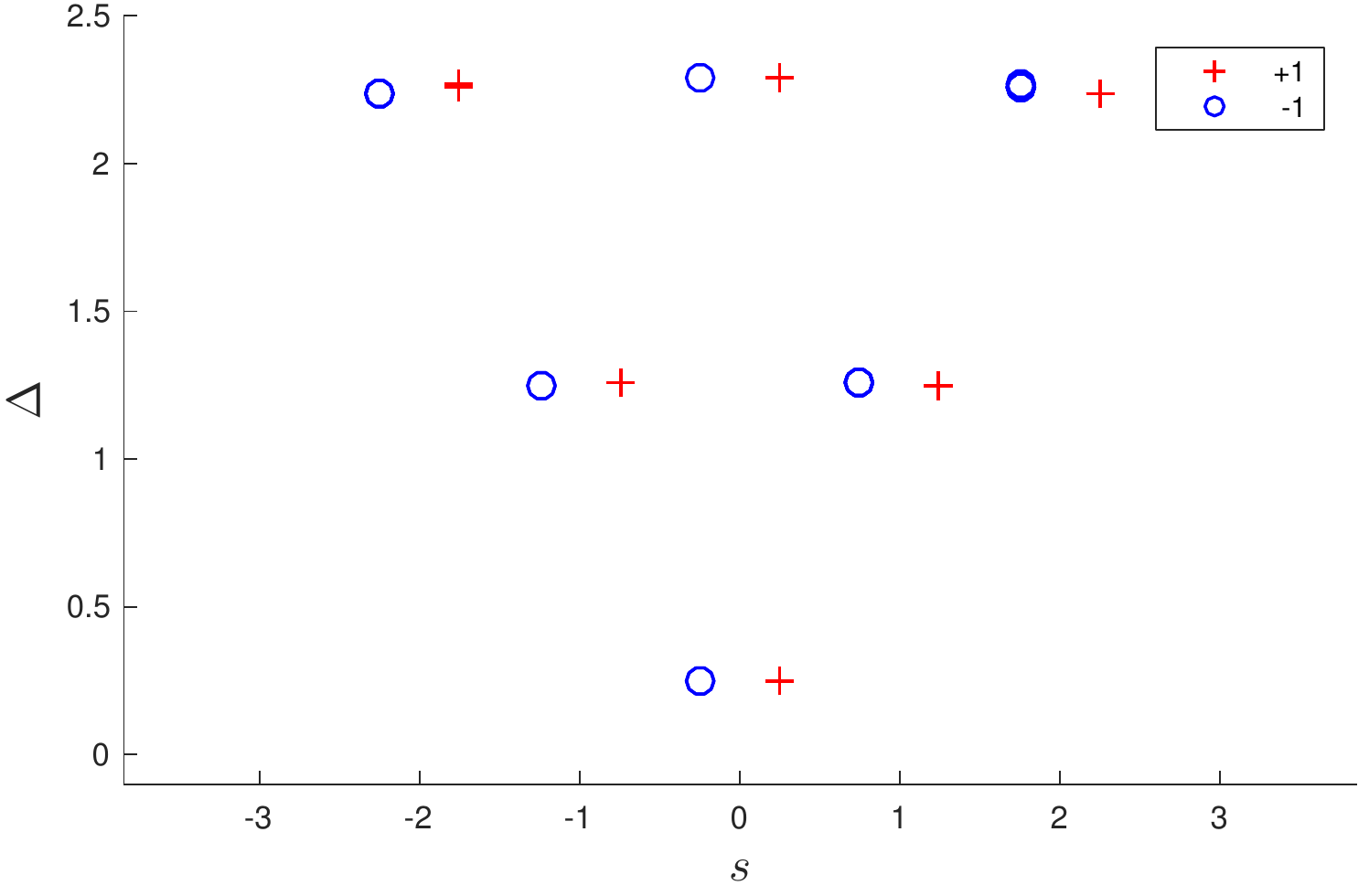}\\ c)
\includegraphics[width=0.22\textwidth]{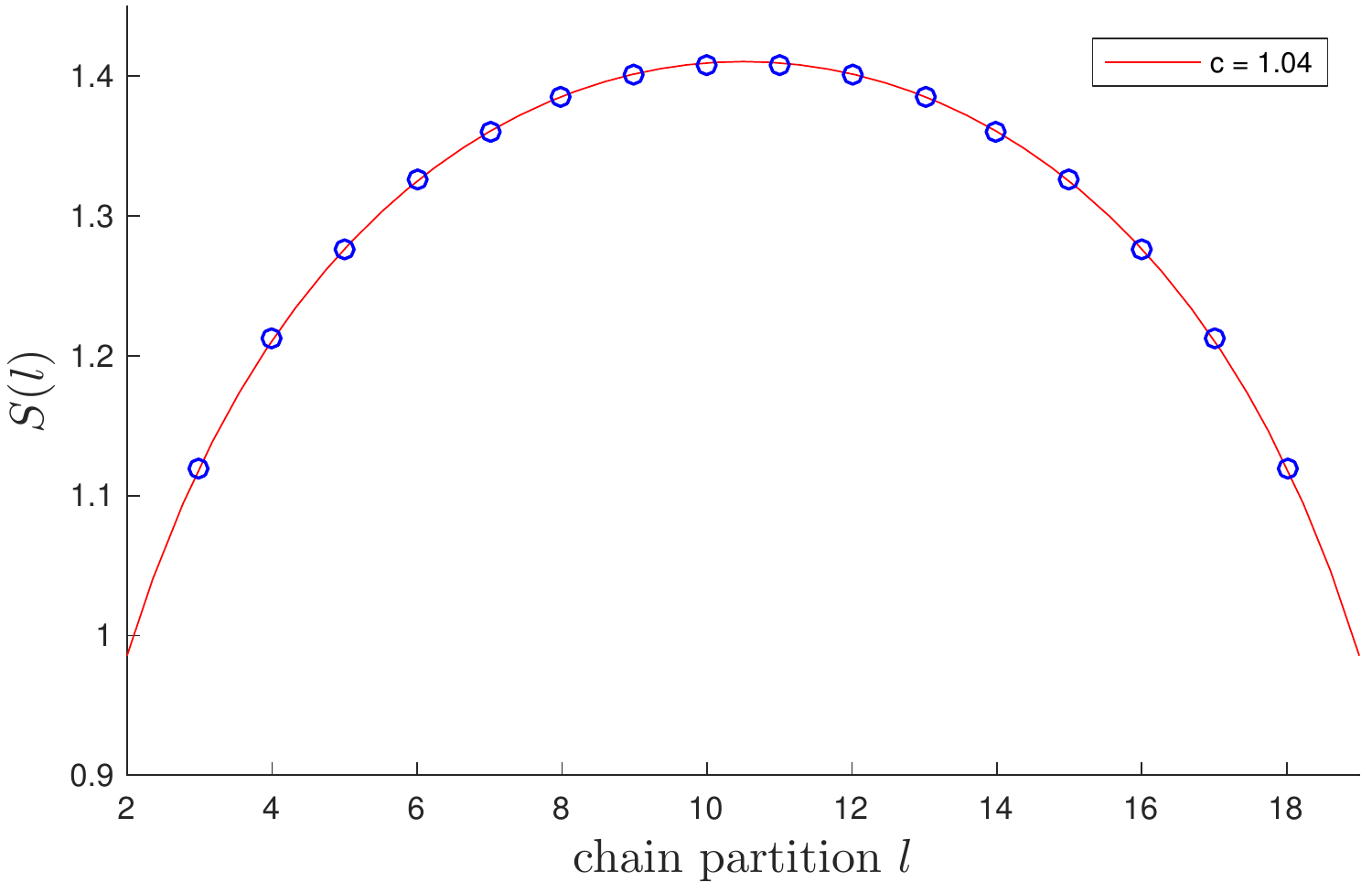}
\caption{(a): Spectrum of $H_{\text{CFT}}$, with $\text{exp}(-H_{\text{CFT}})$ the $\mathbb{Z}_2$  SPT strange correlator partition function transfer matrix. Circumference $L=18$ and ABC branching structure~\cite{ScaffidiRingel}. $\mathbb{Z}_2$ symmetry labels are denoted with $+$ or $o$. The scaling dimension $\Delta=h+\bar{h}$ is plotted against the spin $s=h-\bar{h}$. (b): Spectrum of the same transfer matrix, but with $\mathbb{Z}_2$ twisted boundary conditions. (c): Entanglement entropy $S(l)$ of an interval of length $l$ in the transfer matrix fixed point with $L=21$.}\label{fig:SCZ2}
\end{figure}

\emph{Strange correlator spectra --} To illustrate that CFTs associated to two-dimensional SPTs indeed have 3-cocycles $\alpha(g,h,k)$ in a non-trivial class of $H^3(G,U(1))$ we first employ the strange correlator method \cite{YouBi}. We consider the overlap of the SPT tensor network states on the hexagonal lattice studied in Ref.~\cite{SPTpaper} (which are derived from the discrete path integrals in Ref.~\cite{ChenGu}) with the symmetric product state $\sum_{g\in G} |g\rangle$ on all physical indices, and interpret this overlap as a partition function. The tensor network states of Ref.~\cite{SPTpaper}, and therefore the corresponding partition functions, are completely determined by a group $G$, a 3-cocycle $\alpha(g,h,k)$, and a branching structure. To obtain a finite size CFT spectrum we put the partition function on a cylinder and diagonalize the transfer matrix, which is identified with $\text{exp}(-H_{\text{CFT}})$. The result for $G=\mathbb{Z}_2$ and $\alpha(g,h,k)$ in the non-trivial class of $H^3(\mathbb{Z}_2,U(1))=\mathbb{Z}_2=\{0,1\}$ (we denote the group action as addition mod 2) is shown in Figure~\ref{fig:SCZ2}(a). We rescaled the spectra such that the lowest eigenvalues can be identified with the scaling dimensions $\Delta=h+\bar{h}$ and the momenta with the spins $s=h-\bar{h}$ of the underlying CFT \cite{Affleck,BloteCardy}. The lowest eigenvalues follow the pattern $\Delta=e^2/R^2+m^2R^2/4$, with $e,m\in\mathbb{Z}$, of the free boson CFT at the self-dual radius $R=\sqrt{2}$. From the entanglement entropy scaling of the transfer matrix fixed point\cite{CalabreseCardy} in Figure~\ref{fig:SCZ2}(c) we also find a central charge very close to one. The important information to observe the $\mathbb{Z}_2$ anomaly in this theory is contained in the symmetry labels in the left upper plot of Figure~\ref{fig:SCZ2}. These labels can be obtained via the MPO symmetry action of the original tensor network states as derived in Refs. \cite{ChenLiu,SPTpaper}. The $\mathbb{Z}_2$ symmetry quantum numbers for the compactified boson primaries labeled by charge $e$ and winding $m$ are $(-1)^{e+m}$, in accordance with Ref.~\cite{ChenWen}. From this we deduce that the CFT partition function with boundary conditions twisted in the time direction by the non-trivial element of $\mathbb{Z}_2$ corresponds to $Z_{1|0}=|\chi_1|^2-|\chi_s|^2$, where $\chi_1$ is the character corresponding to the identity primary field, and $\chi_s$ is the character corresponding to the semion primary field. Now we can use the $S$ matrix of the self-dual boson CFT, given by $\frac{1}{\sqrt{2}}\left(\begin{matrix}1&1\\1&-1\end{matrix}\right)$, to find that the partition function with boundary conditions twisted in the spatial direction is $Z_{0|1}=\chi_1\bar{\chi}_s + \chi_s\bar{\chi}_1$. We verified this expression numerically by explicity diagonalizing the transfer matrix with twisted boundary conditions, which we constructed using the MPO techniques from Ref.~\cite{SPTpaper}. The result is shown in Figure \ref{fig:SCZ2}(b). Note that in the twisted spectrum the spins get shifted by $1/4$ \cite{SantosWang}, which is also the shift in the topological spin of symmetry defects in the two-dimensional bulk characteristic for the non-trivial SPT order. Using the results of Ref.~\cite{FrohlichFuchs2}, we can conclude from the spatially twisted partition function that the $\mathbb{Z}_2$ defect line coincides with the semion topological defect. Looking at the $F$-symbols of the semion modular tensor category, we indeed find that recoupling the $\mathbb{Z}_2$ defect gives phases in the non-trivial class of $H^3(\mathbb{Z}_2,U(1))$.

The free boson CFT at the self-dual radius is equivalent to the $SU(2)_1$ Wess-Zumino-Witten CFT. From the twisted partition function $Z_{1|0}=|\chi_1|^2-|\chi_s|^2$ we see that the $\mathbb{Z}_2$ symmetry action is the same as the $g\leftrightarrow -g$ symmetry in the $SU(2)_{k=1}$ CFT, which has been known for some time to be anomalous when the level $k$ is odd \cite{GepnerWitten}. Interestingly, this anomaly has recently been connected to the Lieb-Schultz-Mattis theorem for one-dimensional spin chains \cite{FuruyaOshikawa,JianBi,ChoHsieh,MetlitskiThorngren}. Our analysis shows that the anomaly associated to non-trivial $\mathbb{Z}_2$ SPT phases is the same as the one underlying the Lieb-Schultz-Mattis theorem.

\begin{figure}[h]
a)
\includegraphics[width=0.2\textwidth]{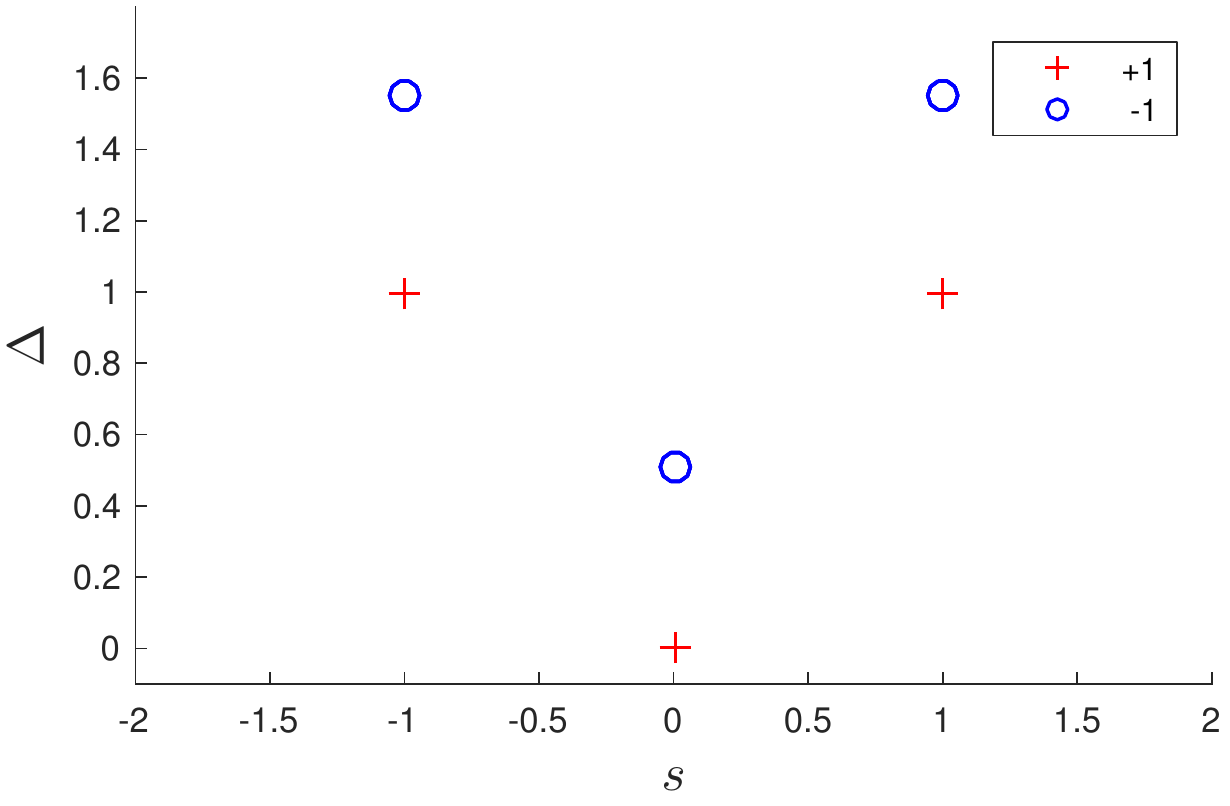}  \hspace{1 mm} b)
\includegraphics[width=0.2\textwidth]{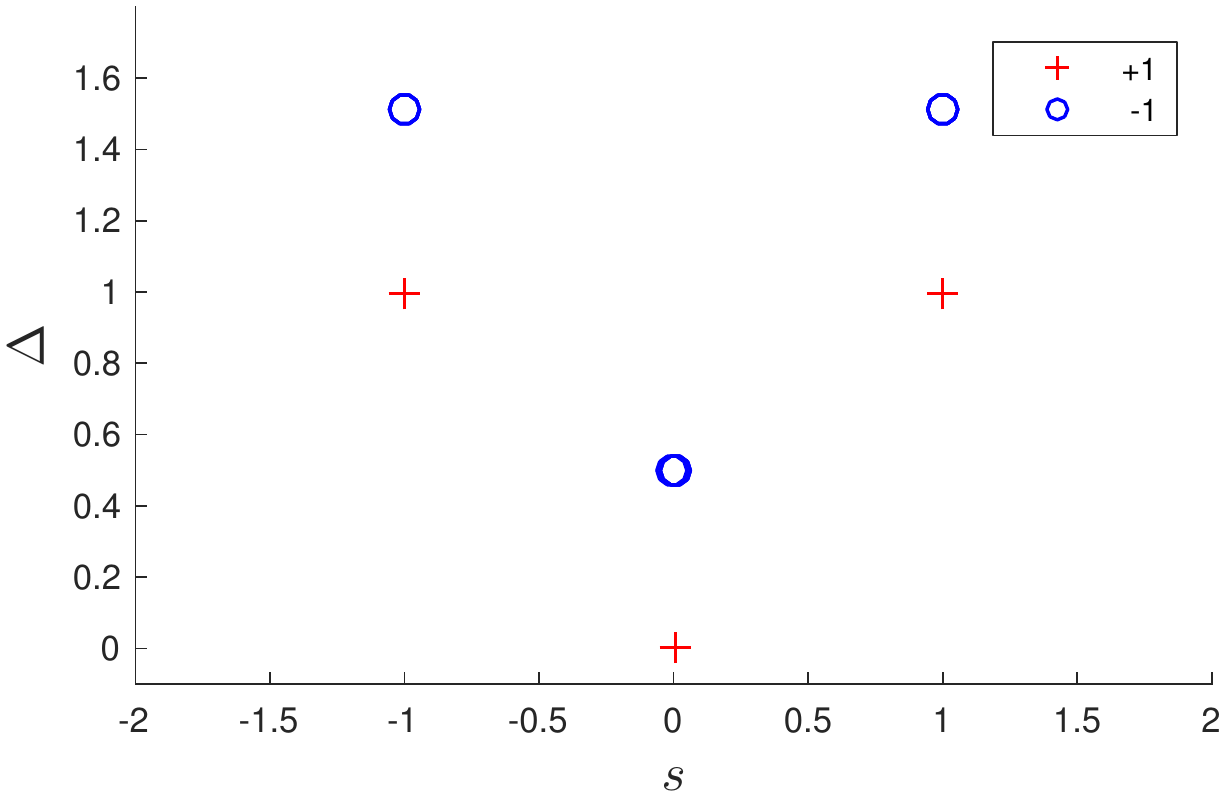} \\ c)
\includegraphics[width=0.2\textwidth]{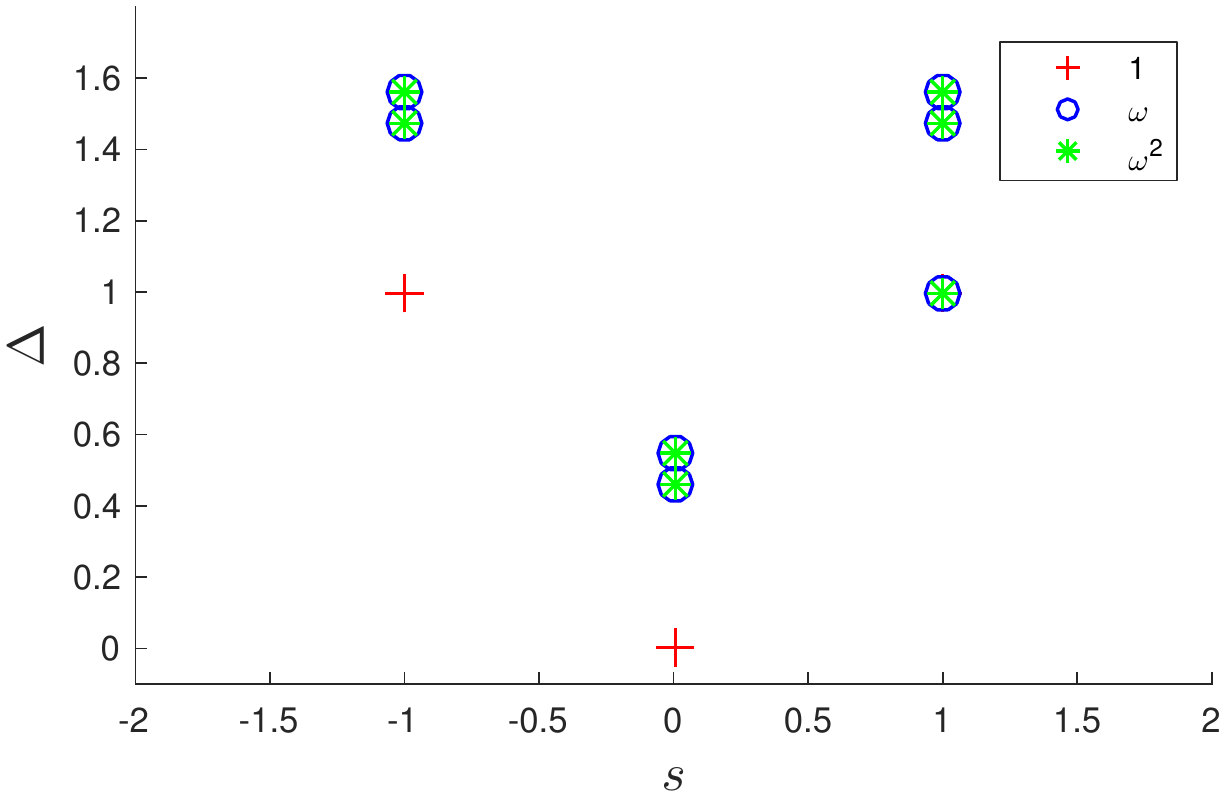}\hspace{1 mm}  d)
\includegraphics[width=0.22\textwidth]{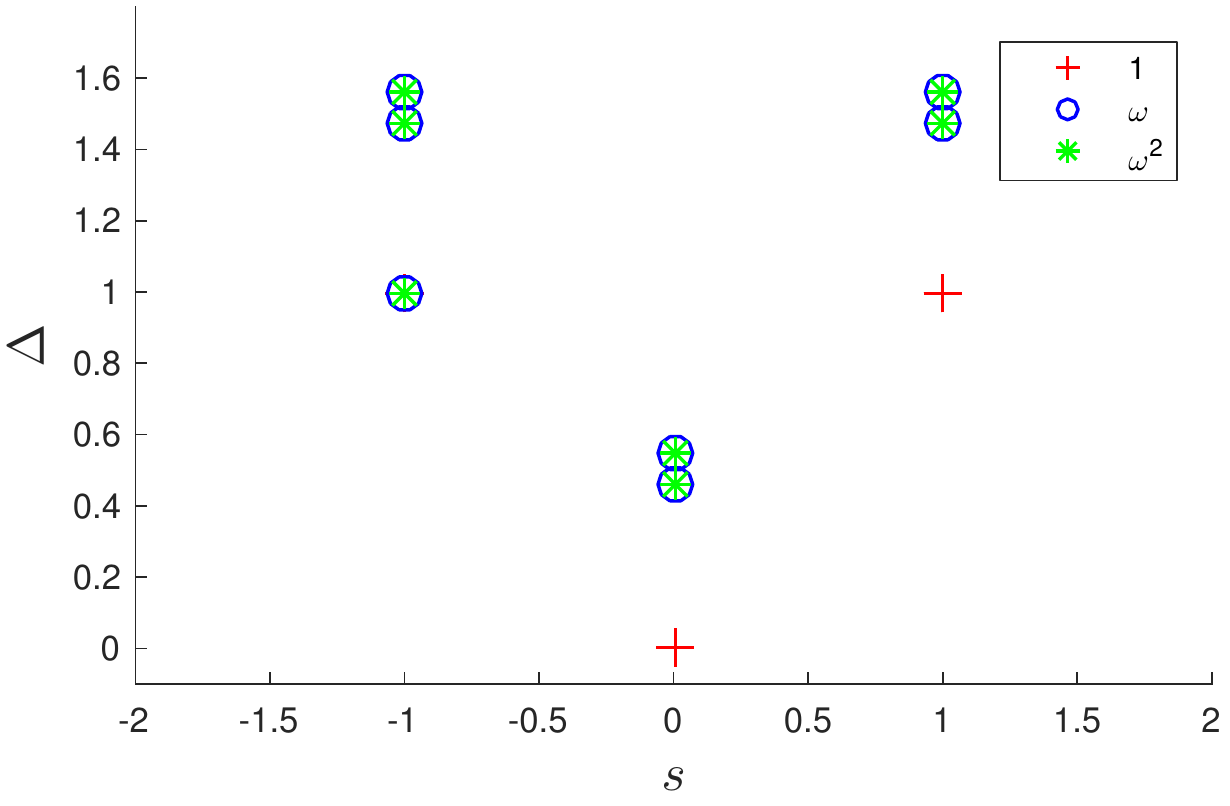}
\caption{Strange correlator spectra obtained with ABC branching structure \cite{ScaffidiRingel} (a): $\mathbb{Z}_2\times\mathbb{Z}_2$ SPT with non-trivial Type II cocycle. L=9. (1,1) symmetry quantum numbers are indicated with $+$ or $o$. (b): $\mathbb{Z}_2\times\mathbb{Z}_2\times \mathbb{Z}_2$ with non-trivial Type III coycle. L=5. (1,1,1) symmetry quantum numbers are shown. (c) and (d): $\mathbb{Z}_3$ with 3-cocycles from both nontrivial
classes in $H^3(\mathbb{Z}_3,U(1))$. L=12. $\omega = \exp(2i\pi /3)$ is the $\mathbb{Z}_3$ quantum number.}\label{fig:SCUntwisted}
\end{figure}

In Figure \ref{fig:SCUntwisted} we show similar spectra for the groups $\mathbb{Z}_2\times\mathbb{Z}_2$ with non-trivial Type II cocycle, $\mathbb{Z}_2\times\mathbb{Z}_2\times\mathbb{Z}_2$ with a non-trivial Type III cocycle, and $\mathbb{Z}_3$ with 3-cocycles from both non-trivial classes in $H^3(\mathbb{Z}_3,U(1))=\mathbb{Z}_3$. For explicit expressions for these 3-cocycles we refer to \cite{Propitius,Zaletel}. The spectra for $\mathbb{Z}_2\times\mathbb{Z}_2$ and $\mathbb{Z}_2\times\mathbb{Z}_2\times\mathbb{Z}_2$ again correspond to a compactified boson with $R=\sqrt{2}$ and the symmetry labels, which respectively denote the eigenvalues of the group elements $(1,1)$ and $(1,1,1)$, are identical to those of $\mathbb{Z}_2$ in Figure \ref{fig:SCZ2}. This implies that both anomalies associated to a non-trivial Type II and Type III cocycle are signaled by a shift of $1/4$ in the topological spins of the corresponding twisted sectors. We note that similar spectra were obtained by diagonalizing one-dimensional lattice Hamiltonians in \cite{Bridgeman2}. For $\mathbb{Z}_3$ we obtain a $R\approx \sqrt{1.85}$ free boson CFT, similar to Ref.~\cite{ScaffidiRingel}. The symmetry labels agree with the general formula $\text{exp}(2i\pi(e+pm)/N)$ for symmetry group $\mathbb{Z}_N$ and a 3-cocycle labeled by $p\in \{0,1,\dots ,N\}= \mathbb{Z}_N=H^3(\mathbb{Z}_N,U(1))$, as proposed in Ref.~\cite{ChenWen}. 

\begin{figure}[t]
a)
\includegraphics[width=0.2\textwidth]{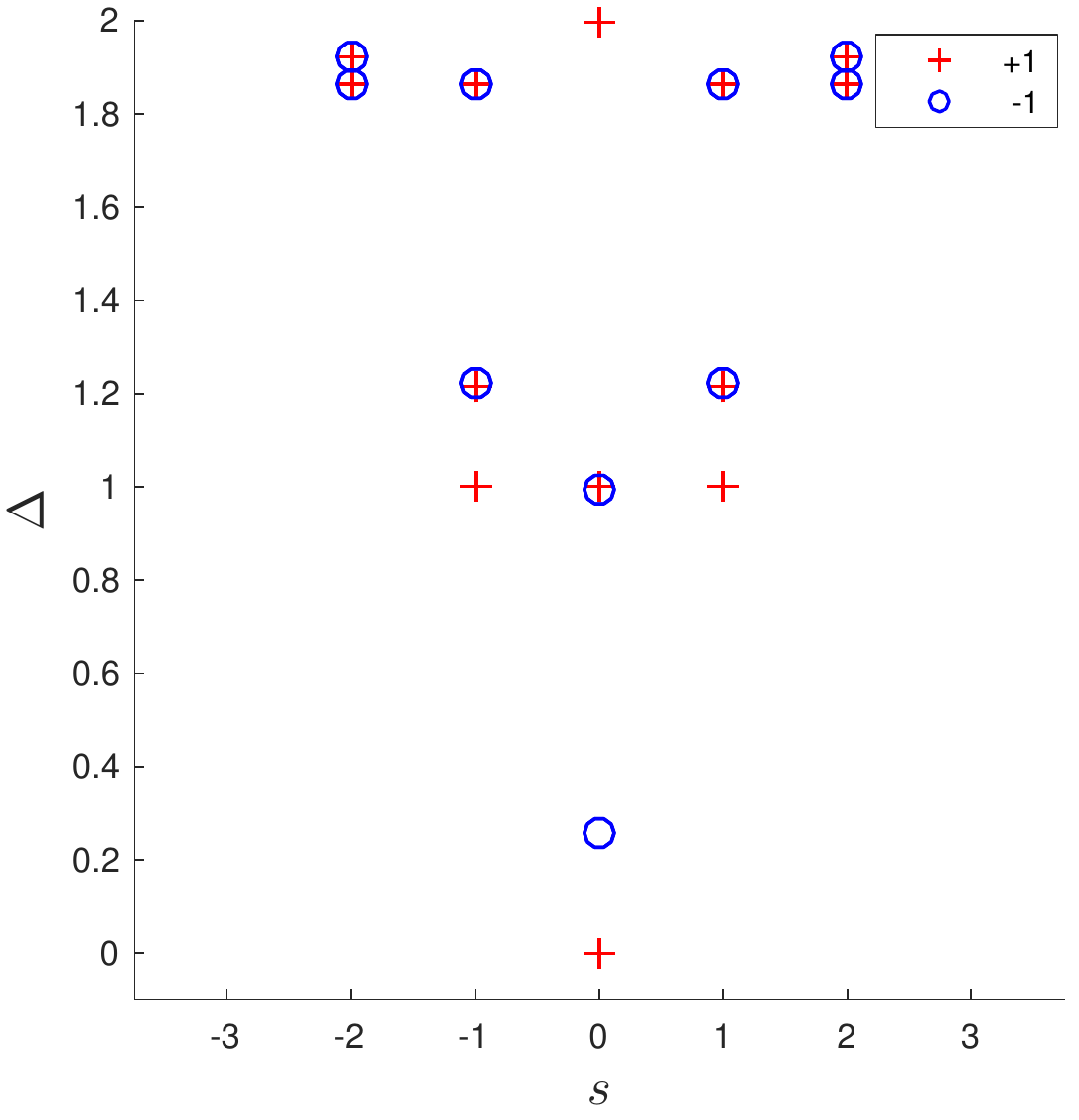} b)
\includegraphics[width=0.2\textwidth]{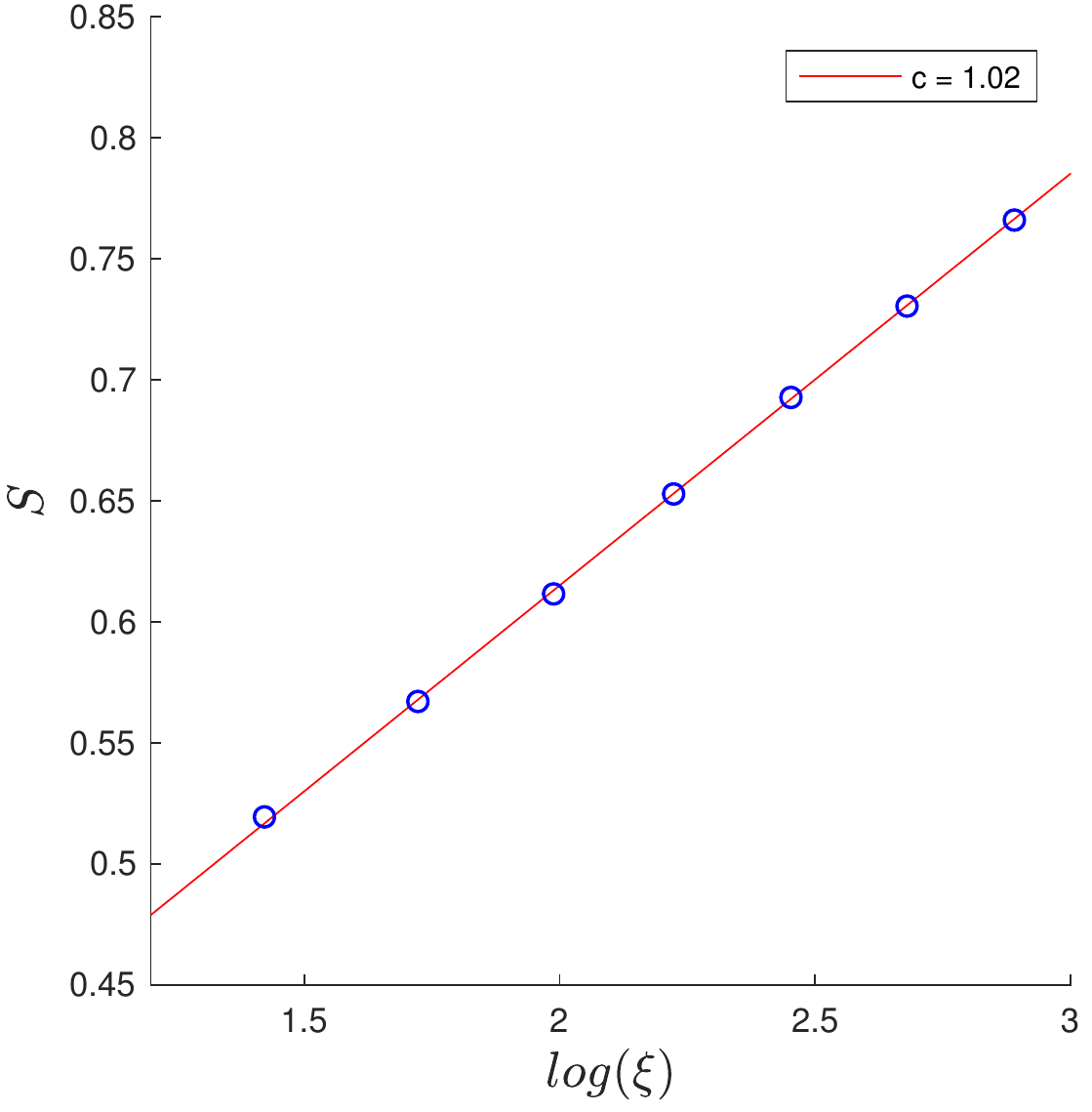}
\caption{(a): Entanglement spectrum for the non-trivial $\mathbb{Z}_2$ SPT with $\mathbb{Z}_2$ symmetry labels. $L=8$, $\lambda = 0.05$ and branching structure as used in Ref.\cite{FermionicPEPSpaper}. (b): Finite-entanglement scaling: Entanglement entropy of a half-infinite partition of the entanglement Hamiltonian ground state versus the logarithm of the correlation length induced via finite bond dimension. The central charge is extracted using $S=c/6\log(\xi)$ \cite{CalabreseCardy}}. \label{fig:Z2ES}
\end{figure}

\emph{Entanglement spectrum --} For the non-trivial $\mathbb{Z}_2$ SPT we also computed the entanglement spectrum. From the same SPT tensor network, now perturbed with the local filtering $\text{exp}(\lambda \sigma_x)$ on the physical indices, we obtain the entanglement Hamiltonian via the prescription of Ref.~\cite{CiracPoilblanc}. The spectrum of the entanglement Hamiltonian on a finite ring is plotted in \ref{fig:Z2ES}(a). We again recognize a compactified free boson CFT, but now with $R=2$, which is again rational. In Figure \ref{fig:Z2ES}(b) we extract the central charge $c$ for the entanglement Hamiltonian in the thermodynamic limit, using finite-entanglement scaling \cite{Tagliacozzo,PollmannMukerjee}. Again we find $c$ to lie very close to one. To label the strange correlator spectra with symmetry quantum numbers, we used the explicit MPO constructions from Refs.~\cite{ChenLiu,SPTpaper}. This is of course not feasible for the numerical detection of SPT order in generic many-body ground states. The advantage of the entanglement spectrum, however, is that one does not need to know the MPOs to find the symmetry quantum numbers. We put the two-dimensional system on an infinite cylinder, and define a modified reduced density matrix $\rho_L^g$ for the half-infinite left part as $\rho_L^g\equiv \text{tr}_R\left((\mathds{1}_L\otimes U(g)_R)|\psi\rangle\langle\psi|\right)$, where $|\psi\rangle$ is the ground state and $U(g)_R$ with $g\in G$ is the on-site global symmetry action on the right half. We can now project the reduced density matrix on the subspace corresponding to irrep $\mu$ with character $r^\mu(g)$ using $\sum_{g\in G} r^{\mu}(g)\rho^g_L$. The result of this procedure for $\mathbb{Z}_2$ is shown in Figure~\ref{fig:Z2ES}. Again the $\mathbb{Z}_2$ quantum numbers are given by $(-1)^{e+m}$. For the $R=2$ free boson CFT we cannot associate a primary field with the $\mathbb{Z}_2$ defect, since it only commutes with the Virasoro algebra, and not with the full extended chiral algebra. Using Poisson resummation one can still perform the $S$ transformation to obtain

\begin{align}\label{eq:poisson}
\frac{1}{\eta(-1/\tau)\bar{\eta}(-1/\bar{\tau})}\sum_{e,m\in\mathbb{Z}}(-1)^{e+m}\tilde{q}^{\frac{1}{2}\left(\frac{e}{R}+m\frac{R}{2}\right)^2}\bar{\tilde{q}}^{\frac{1}{2}\left(\frac{e}{R}-m\frac{R}{2}\right)^2} \nonumber \\
= \frac{1}{\eta(\tau)\bar{\eta}(\bar{\tau})}\sum_{e,m\in\mathbb{Z}+1/2}q^{\frac{1}{2}\left(\frac{e}{R}+m\frac{R}{2}\right)^2}\bar{q}^{\frac{1}{2}\left(\frac{e}{R}-m\frac{R}{2}\right)^2}\, ,
\end{align}
with $q=\text{exp}(2\pi i \tau)$ and $\tilde{q}=\text{exp}(-2\pi i/ \tau)$. From \eqref{eq:poisson} we again see the characteristic shift of $1/4$ in the spins of the twisted spectrum. Equation \eqref{eq:poisson} can easily be generalized for all $\mathbb{Z}_N$ SPTs, showing that the formula from Ref.~\cite{ChenWen} for the symmetry labels in the untwisted sector and the formula from Ref.~\cite{SantosWang} for the scaling dimensions and spins in the twisted sector are equivalent. So to numerically detect the SPT order one only needs a single, untwisted ground state. However, there might be situations where the CFT techniques do not apply, in which case one explicitly has to compute the twisted ground states, as was done in Ref.~\cite{Zaletel}.

\emph{Outlook --} In this work we restricted ourselves to unitary, on-site symmetries in bosonic systems. For the anomaly analysis of CFTs describing edge theories of phases protected by time-reversal or spatial reflection it is known that one needs to consider partition functions on unorientable spacetimes \cite{HsiehSule}. It would be interesting to connect this to the approach of this paper. Using the recent results of Refs. \cite{WilliamsonBultinck,FermionicPEPSpaper,AasenLake} the defect line formalism can be extended to fermionic CFTs, where an objective would be to make connection with the the results of Ref. \cite{FreedVafa} on global anomalies in orbifolds. The arguments based on defect lines can also be generalized to higher dimensional systems. This is in contrast to most other CFT techniques, which rely on the analytical properties of conformal symmetry in one spatial dimension. We leave these directions open for future work.

\begin{acknowledgements}
We thank Dominic J. Williamson for fruitful collaborations on related work and for pointing out Ref.~\cite{Mignard} to us, and Zohar Ringel for explaining some technical details from Ref.~\cite{ScaffidiRingel}. We ackowledge helpful conversations with Jacob Bridgeman and David Aasen. This work was supported by the Austrian Science Fund (FWF) through grants ViCoM and FoQuS, and the EC through the grants QUTE (647905) and ERQUAF (715861). J.H. acknowledges the support from the Research Foundation Flanders (FWO).
\end{acknowledgements}

\bibliography{SPTDetection.bib}

\pagebreak
\appendix
\vspace{5 mm}
\begin{center}
\large{\textbf{Supplementary material}}
\end{center}

Here we study the properties of group defect lines in more detail and derive the orbifold properties under the modular transformations. To start, we first note that the group defect lines satisfy following property:

\begin{align}\label{inverse}
\includegraphics[width=0.2\textwidth]{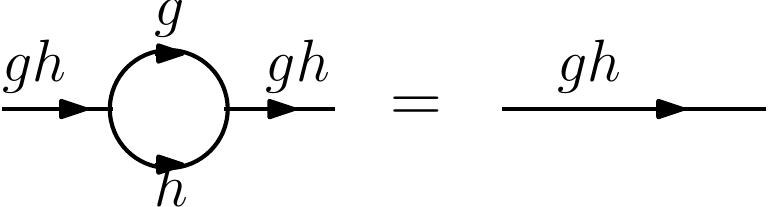}\, ,
\end{align}
which we will use multiple times below. Let us now also introduce following notation for the vertex operator $V_{(g,g^{-1})}$:

\begin{align}
\includegraphics[width=0.22\textwidth]{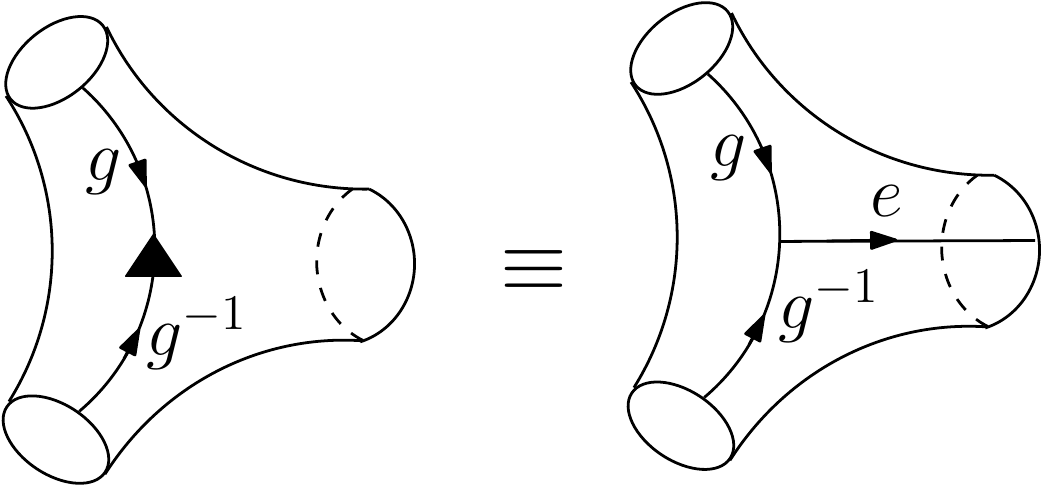}
\end{align}
From \eqref{inverse} it then follows that 

\begin{align}
\includegraphics[width=0.05\textwidth]{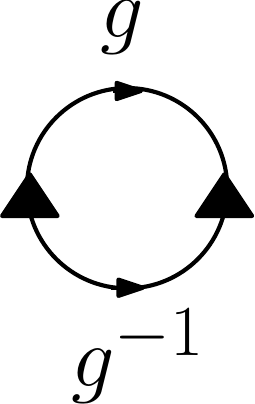}
\end{align}
is equivalent to the empty graph, i.e. no defect line. This implies following identity:

\begin{align}\label{s1}
\includegraphics[width=0.18\textwidth]{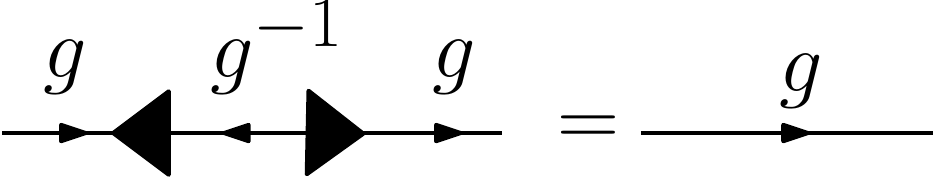}\, .
\end{align}
Now we can also connect the vertex operators $V_{(g,g^{-1})}$ and $V_{(g^{-1},g)}$ and obtain

\begin{align}\label{s2}
\includegraphics[width=0.19\textwidth]{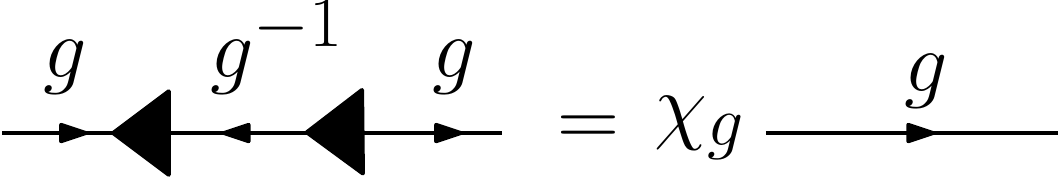}\, ,
\end{align}
where $\chi_g$ is a phase. Using the property shown in Fig. \ref{fig:vertex} one can easily derive that $\chi_g = \alpha(g,g^{-1},g)$. From \eqref{s1} and \eqref{s2} we can conclude that

\begin{align}
\includegraphics[width=0.19\textwidth]{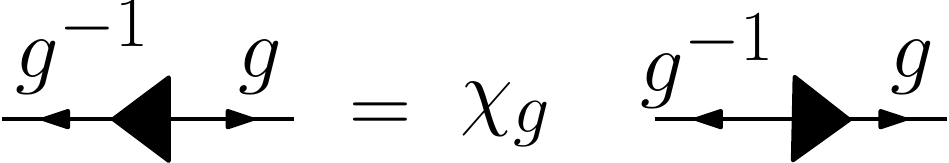}\, .
\end{align}
To derive the modular transformation of the orbifold partition function we also need following two identities:

\begin{align}\label{s3}
\includegraphics[width=0.29\textwidth]{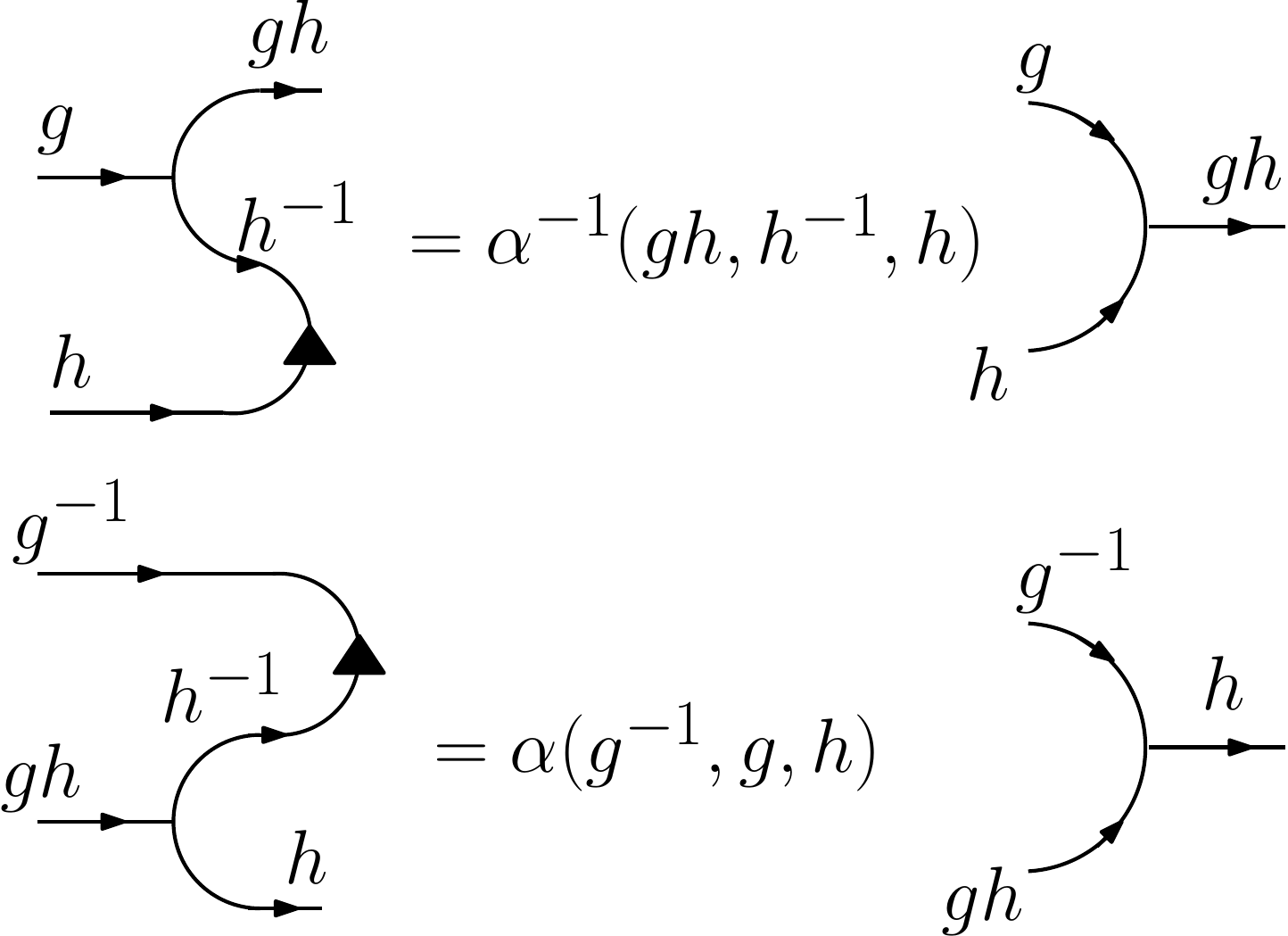}
\end{align}
which can again be derived using Fig. \ref{fig:vertex} in the following way:

\begin{align}
\includegraphics[width=0.36\textwidth]{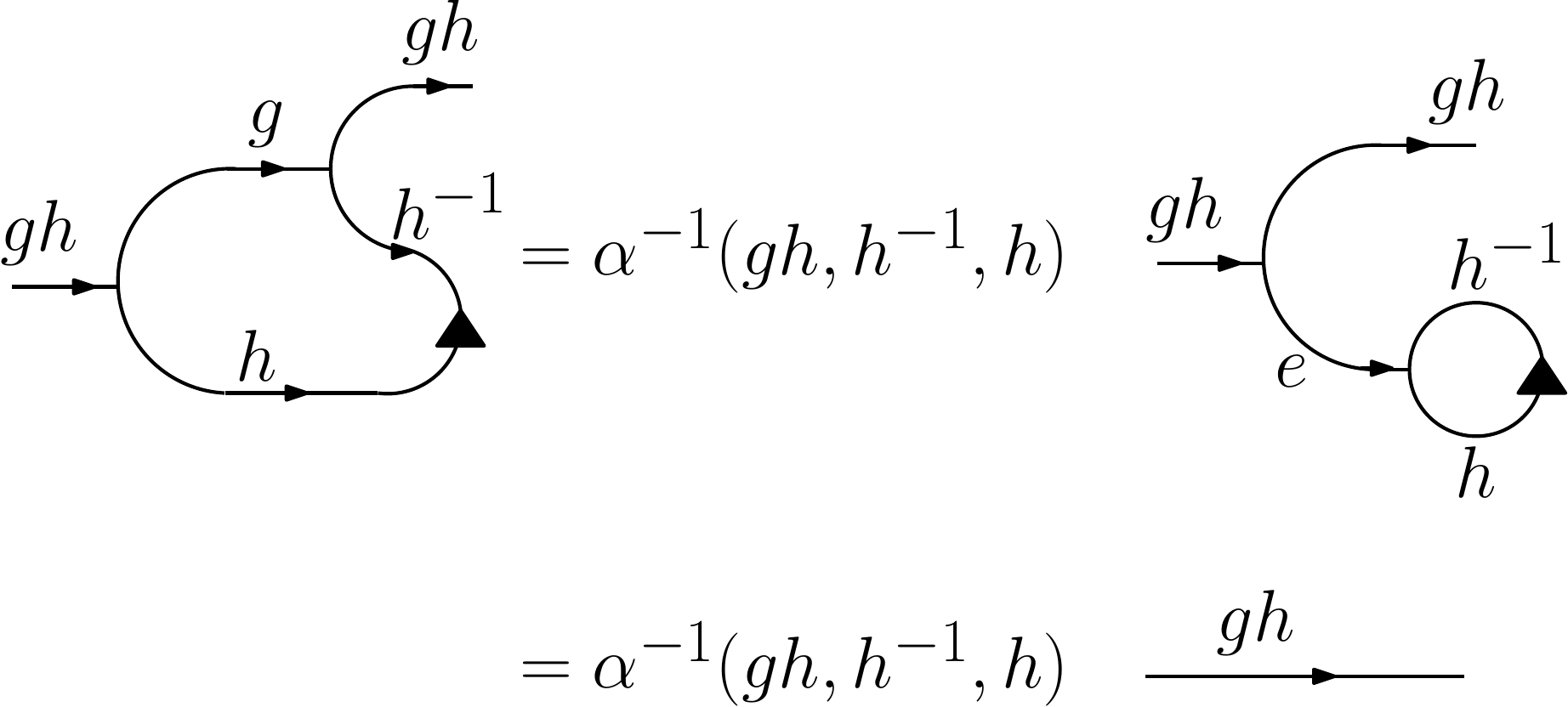}\, ,
\end{align}
and similarly for the second identity in \eqref{s3}.

Now we are ready to derive the modular properties of the orbifold partition function. Let us first start with the $T$ transformation, which is defined as

\begin{align}
\includegraphics[width=0.3\textwidth]{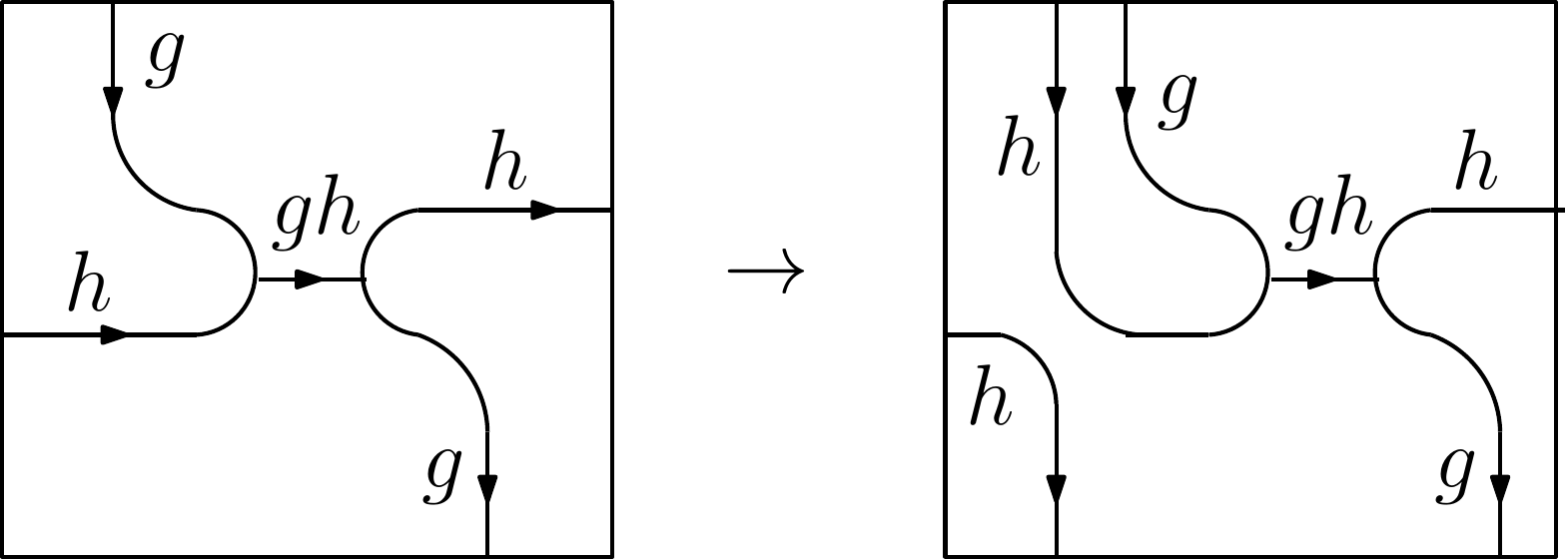}\, .
\end{align}
Using the relations above we can now obtain

\begin{align}
\includegraphics[width=0.35\textwidth]{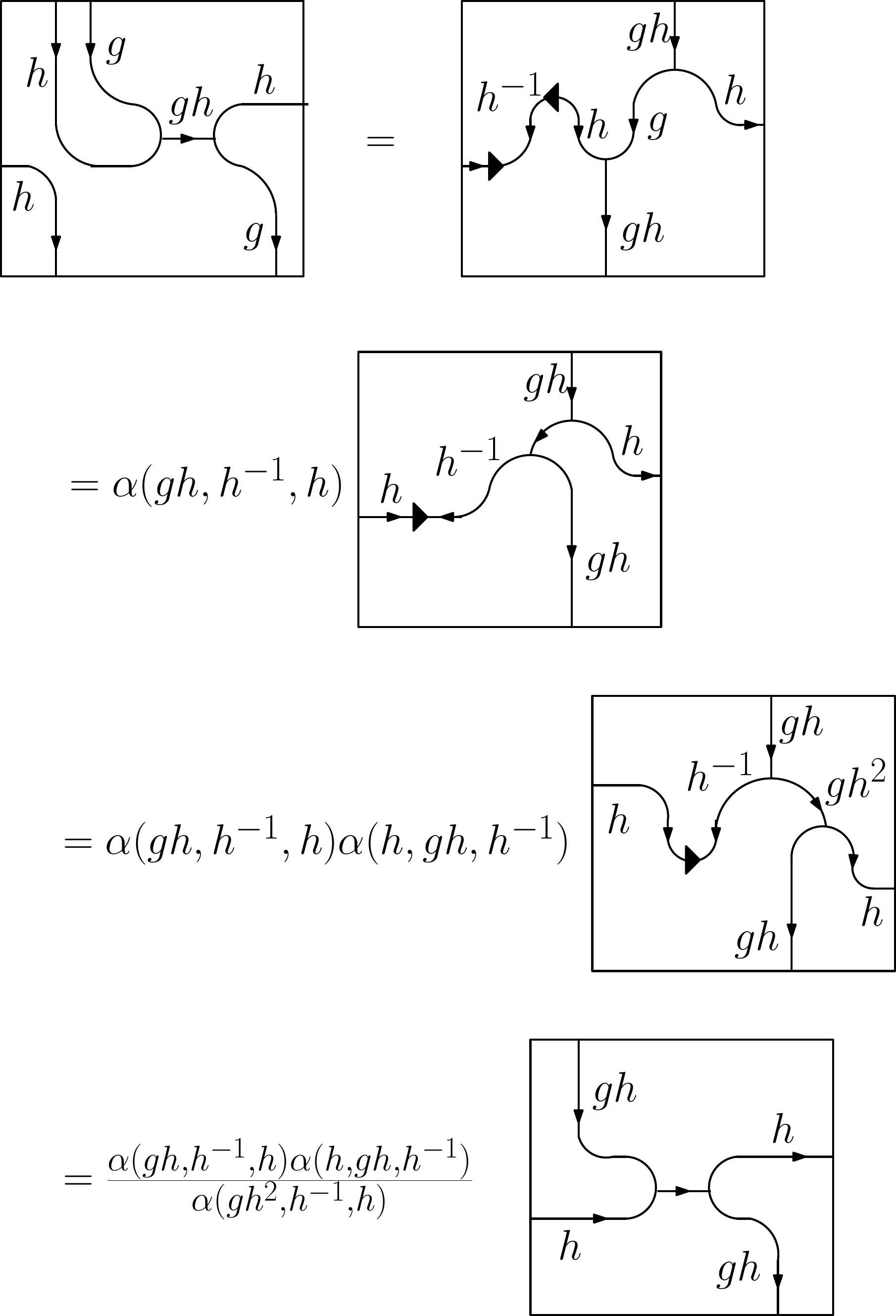}\, ,
\end{align}
which gives \eqref{eq:T} after a few applications of the 3-cocycle relation. The $S$ transformation correponds to

\pagebreak

\begin{align}
\includegraphics[width=0.32\textwidth]{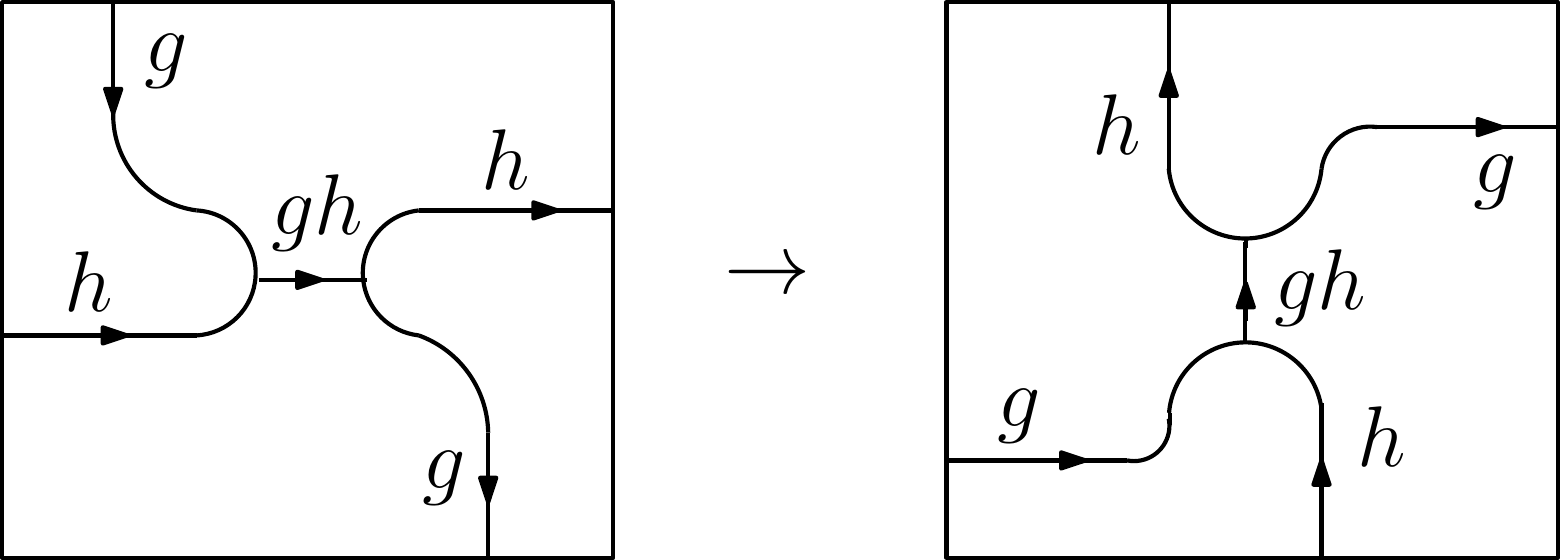}\, .
\end{align}
Similar to the $T$ matrix we can now do a series of manipulations to obtain

\begin{align}
\includegraphics[width=0.35\textwidth]{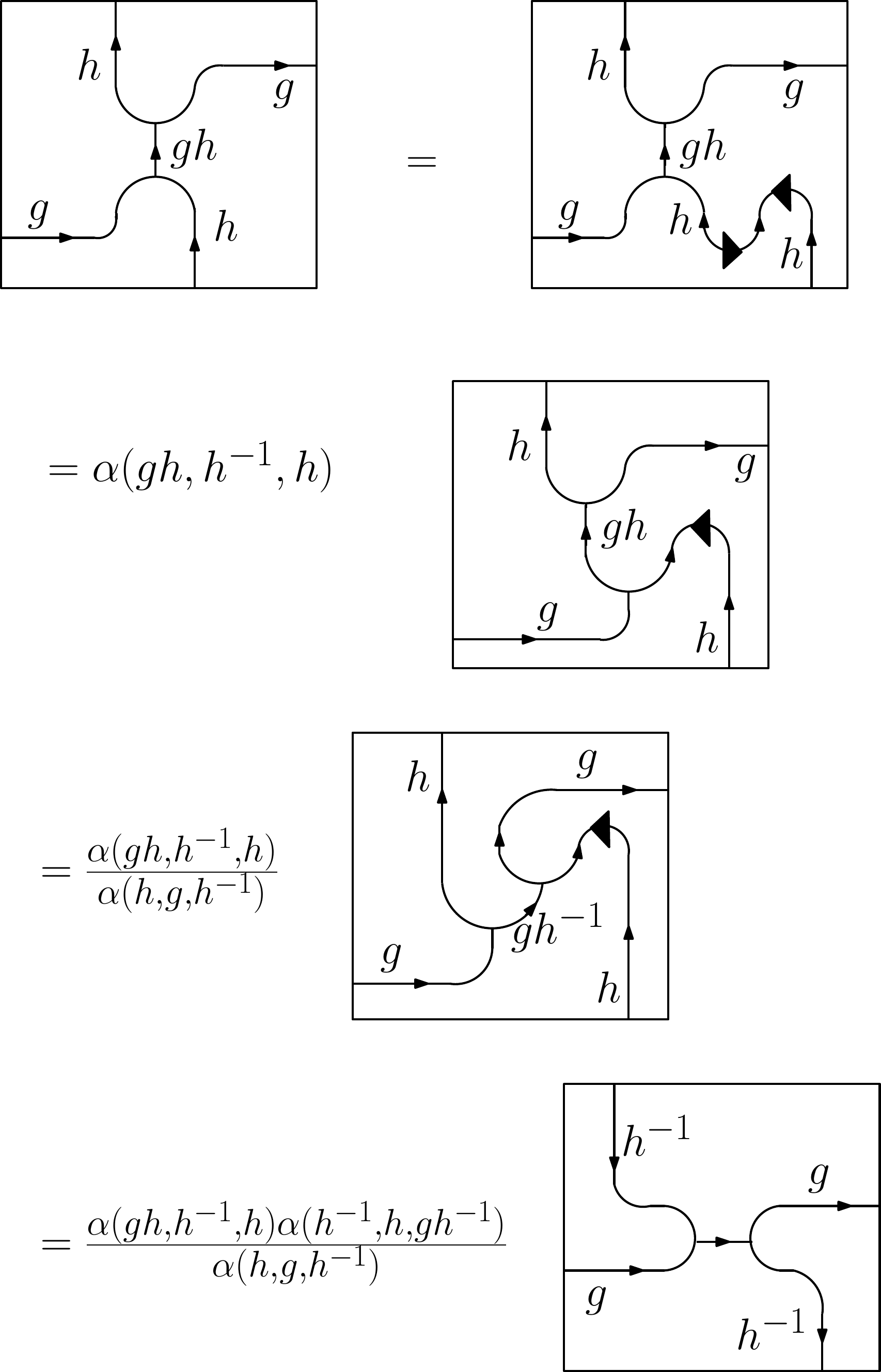}\, ,
\end{align}
which, combined with the 3-cocycle relation, gives equation \eqref{eq:S}.

\end{document}